\documentclass[twoside,twocolumn,9pt]{article}
\usepackage{amssymb}
\usepackage{extsizes}
\usepackage[super,sort&compress,comma]{natbib} 
\usepackage[version=3]{mhchem}
\usepackage[left=1.5cm, right=1.5cm, top=1.785cm, bottom=2.0cm]{geometry}
\usepackage{balance}
\usepackage{mathptmx}
\usepackage{sectsty}
\usepackage{graphicx} 
\usepackage{lastpage}
\usepackage[format=plain,justification=justified,singlelinecheck=false,font={stretch=1.125,small,sf},labelfont=bf,labelsep=space]{caption}
\usepackage{float}
\usepackage{fancyhdr}
\usepackage{fnpos}
\usepackage[english]{babel}
\usepackage{array}
\usepackage{droidsans}
\usepackage{charter}
\usepackage[T1]{fontenc}
\usepackage[usenames,dvipsnames]{xcolor}
\usepackage{setspace}
\usepackage[compact]{titlesec}
\usepackage{hyperref}

\usepackage{epstopdf}

\definecolor{cream}{RGB}{222,217,201}

\begin{document}

\pagestyle{fancy}
\thispagestyle{plain}
\fancypagestyle{plain}{
\renewcommand{\headrulewidth}{0pt}
}

\makeFNbottom
\makeatletter
\renewcommand\LARGE{\@setfontsize\LARGE{15pt}{17}}
\renewcommand\Large{\@setfontsize\Large{12pt}{14}}
\renewcommand\large{\@setfontsize\large{10pt}{12}}
\renewcommand\footnotesize{\@setfontsize\footnotesize{7pt}{10}}
\makeatother

\renewcommand{\thefootnote}{\fnsymbol{footnote}}
\renewcommand\footnoterule{\vspace*{1pt}%
\color{cream}\hrule width 3.5in height 0.4pt \color{black}\vspace*{5pt}} 
\setcounter{secnumdepth}{5}

\makeatletter 
\renewcommand\@biblabel[1]{#1}            
\renewcommand\@makefntext[1]%
{\noindent\makebox[0pt][r]{\@thefnmark\,}#1}
\makeatother 
\renewcommand{\figurename}{\small{Fig.}~}
\sectionfont{\sffamily\Large}
\subsectionfont{\normalsize}
\subsubsectionfont{\bf}
\setstretch{1.125} 
\setlength{\skip\footins}{0.8cm}
\setlength{\footnotesep}{0.25cm}
\setlength{\jot}{10pt}
\titlespacing*{\section}{0pt}{4pt}{4pt}
\titlespacing*{\subsection}{0pt}{15pt}{1pt}

\fancyfoot{}
\fancyfoot[LO,RE]{\vspace{-7.1pt}\includegraphics[height=9pt]{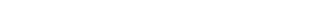}}
\fancyfoot[CO]{\vspace{-7.1pt}\hspace{13.2cm}\includegraphics{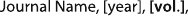}}
\fancyfoot[CE]{\vspace{-7.2pt}\hspace{-14.2cm}\includegraphics{head_foot/RF}}
\fancyfoot[RO]{\footnotesize{\sffamily{1--\pageref{LastPage} ~\textbar  \hspace{2pt}\thepage}}}
\fancyfoot[LE]{\footnotesize{\sffamily{\thepage~\textbar\hspace{3.45cm} 1--\pageref{LastPage}}}}
\fancyhead{}
\renewcommand{\headrulewidth}{0pt} 
\renewcommand{\footrulewidth}{0pt}
\setlength{\arrayrulewidth}{1pt}
\setlength{\columnsep}{6.5mm}
\setlength\bibsep{1pt}

\makeatletter 
\newlength{\figrulesep} 
\setlength{\figrulesep}{0.5\textfloatsep} 

\newcommand{\topfigrule}{\vspace*{-1pt}%
\noindent{\color{cream}\rule[-\figrulesep]{\columnwidth}{1.5pt}} }

\newcommand{\botfigrule}{\vspace*{-2pt}%
\noindent{\color{cream}\rule[\figrulesep]{\columnwidth}{1.5pt}} }

\newcommand{\dblfigrule}{\vspace*{-1pt}%
\noindent{\color{cream}\rule[-\figrulesep]{\textwidth}{1.5pt}} }

\makeatother

\twocolumn[
  \begin{@twocolumnfalse}
{\includegraphics[height=30pt]{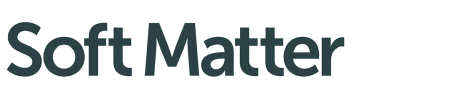}\hfill\raisebox{0pt}[0pt][0pt]{\includegraphics[height=55pt]{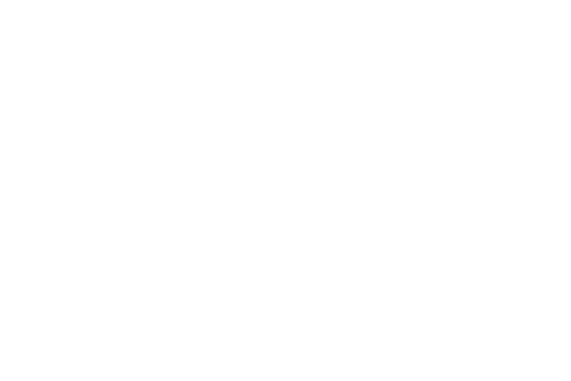}}\\[1ex]
\includegraphics[width=18.5cm]{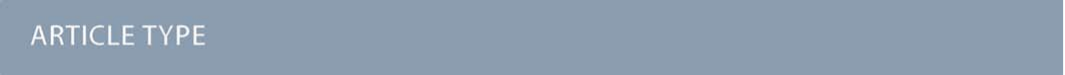}}\par
\vspace{1em}
\sffamily
\begin{tabular}{m{4.5cm} p{13.5cm} }

\includegraphics{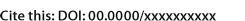} & \noindent\LARGE{\textbf{Symmetry-breaking motility of an active hinge \newline in a crowded channel}} \\
\vspace{0.3cm} & \vspace{0.3cm} \\


 & \noindent\large{Leonardo Garibaldi Rigon\textit{$^{a}$} and Yongjoo Baek$^{\ast}$\textit{$^{a}$}} \\

\includegraphics{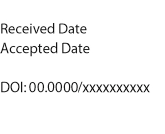} & \noindent\normalsize{A recent experiment [Son \textit{et al., Soft Matter}, 2024, \textbf{20}, 2777-2788] showed that self-propelled particles confined within a circular boundary filled with granular medium spontaneously form a motile cluster that stays on the boundary. This cluster exhibits persistent (counter)clockwise motion driven by symmetry breaking, which arises from a positive feedback between the asymmetry of the cluster and those of the surrounding granular medium. To investigate this symmetry-breaking mechanism in broader contexts, we propose and analyze the dynamics of an active hinge moving through a crowded two-dimensional channel. Through extensive numerical simulations, we find that the lifetime of the hinge’s motile state varies nonmonotonically with both the packing fraction of the granular medium and the strength of self-propulsion. Furthermore, we observe an abrupt transition in the configuration of passive particles that sustain hinge motility as the hinge's maximum angle relative to the channel wall increases. These findings point to the possibility of designing superstructures composed of passive granular media doped with a small number of active elements, whose dynamic modes can be switched by tuning the properties of their components.}

\end{tabular}

 \end{@twocolumnfalse} \vspace{0.6cm}

  ]

\renewcommand*\rmdefault{bch}\normalfont\upshape
\rmfamily
\section*{}
\vspace{-1cm}


\footnotetext{\textit{$^{a}$~Department of Physics and Astronomy \& Center for Theoretical Physics, Seoul National University, Seoul 08826, Korea.}}
\footnotetext{\textit{$^{\ast}$~E-mail: y.baek@snu.ac.kr}}



\section{Introduction}

Active matter encompasses a broad class of nonequilibrium systems in which the constituent particles, called active particles, consume energy to generate motion~\cite{RamaswamyARCMP2010,MarchettiRMP2013,BechingerRMP2016,RamaswamyJSM2017,JulicherRPP2018,GompperJPCM2020,BowickPRX2022,VrugtEPJE2025}. When subjected to spatially asymmetric environments, like a ratchet potential, such active particles produce nonequilibrium currents in the steady state, a phenomenon known as current rectification~\cite{AngelaniNJP2010,KaiserPRL2014,MalloryPRE2014,SmallenburgPRE2015} that has found applications in the design of targeted delivery systems~\cite{KoumakisNComms2013} and self-starting micromotors~\cite{AngelaniPRL2009,SokolovPNAS2010,DiLeonardoPNAS2010}. However, even when the environments are perfectly symmetric, nonequilibrium currents can still emerge through symmetry breaking. This can occur by active particles self-organizing themselves to form asymmetric structures~\cite{TjhungPNAS2012,DeMagistrisSM2014}, by a rigid passive object experiencing a negative drag exerted by the surrounding active particles~\cite{GranekPRL2022,KimPRE2024,WangNSO2024,PeiPRE2025}, or by active particles deforming a soft passive object (like polymer chains) to form an asymmetric structure~\cite{NikolaPRL2016,ShinPCCP2017,LiSM2017,AporvariSoftMatter2020}.

Meanwhile, vibrated granular particles form an important subclass of active matter. In addition to possessing both liquid-like and solid-like properties like other granular systems~\cite{melby2005dynamics}, they are known to exhibit collective motion~\cite{yamada2003coherent,kudrolli2008swarming,deseigne2010collective, deseigne2012vibrated,weber2013long}, crystallization~\cite{briand2018spontaneously}, demixing~\cite{scholz2018rotating}, and nontrivial dynamics of topological defects~\cite{digregorio2022unified}. If such systems consist of a small number of active dopants (``active'' in the sense that they propel themselves in the direction of their polarity) mixed with a majority of passive granular particles ( ``passive'' in the sense that they are isotropic and lack any persistent propulsion direction, despite being possibly athermal), the interplay between the active and passive granular particles have been shown to promote crystallization~\cite{ni2013pushing, kummel2015formation}, facilitate the removal of grain boundaries in polycrystals~\cite{van2016fabricating}, and induce coherent particle transport called ``flocking''~\cite{kumar2014flocking, soni2020phases}.

\begin{figure}
\centering
  \includegraphics[height=2.7cm]{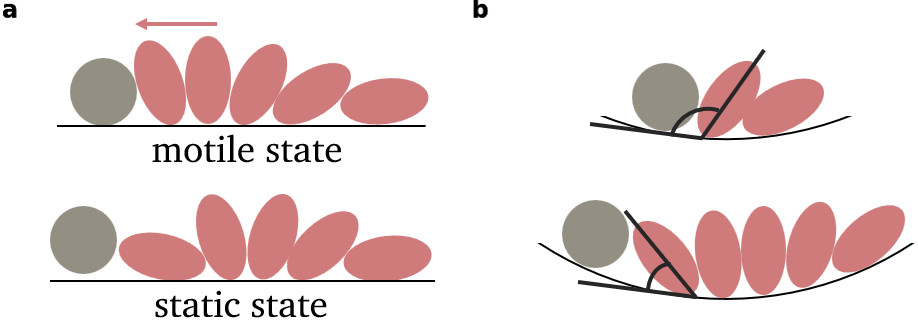}
  \caption{Schematics of the self-propelled particle cluster reported in Son \textit{et al.}~\cite{son2024dynamics} (a) The moving state persists as long as a disk stays stuck in front of the cluster, maintaining its asymmetric shape. Once the disk is removed, the cluster loses the asymmetry and is stuck in a static state. (b) The angle formed by the wall and the foremost particle of the cluster is an important factor determining the lifetime of the motile state.}
  \label{fgr:angle}
\end{figure}

In particular, to observe flocking using a vibrated granular medium in a two-dimensional confinement, it is well known that the confinement boundary has to be shaped like petals~\cite{deseigne2010collective,deseigne2012vibrated} to prevent the accumulation of a large, static cluster of active particles~\cite{kudrolli2008swarming, elgeti2009self, deblais2018boundaries,wensink2008aggregation}. However, a recent experiment~\cite{son2024dynamics} found that when the number of active particles is not too large, they can form a coherent, motile boundary cluster that moves persistently in one direction (clockwise or counterclockwise) even in a perfectly circular confinement lacking petals. This phenomenon can be understood as follows: The motion of the active particle cluster deforms the surrounding passive medium, whose asymmetric structure in turn helps the active particle cluster move persistently in its direction of motion, as illustrated in Fig.~\ref{fgr:angle}. In this sense, the phenomenon can be regarded as an analog of the symmetry-breaking current rectification induced by deformation, where the passive granular medium plays the role of the soft passive object.

When current rectification arises from symmetry-breaking mechanisms, persistence of the resulting motile structures may exhibit nonmonotonic dependence on the parameters describing constituent particles since symmetry breaking may occur only when such parameters fall within limited ranges. This can reveal effective methods for designing superstructures with tunable dynamical properties. In this study, by introducing a simplified model called the active hinge model, we further investigate how the persistence of the symmetry-broken structure is affected by the physical properties of the active and passive granular particles. We find that the cluster’s directional persistence varies nonmonotonically with the packing fraction of the granular medium and the self-propulsion force. Moreover, we observe that the passive particle configuration responsible for the persistence of the motile state exhibits an abrupt change as the maximum angle between the hinge and the channel wall increases, which is a close analog of how the boundary cluster in the experiment of Son \textit{et al.} suddenly loses motility as the number of active particles increases.

The remainder of this paper is organized as follows. In Section \ref{sec:model}, we introduce the model and simulation details. Section \ref{sec:symmetrybreaking} focuses on the emergence of the motile state sustained by a symmetry-breaking mechanism. In Section \ref{sec:results}, we present a detailed analysis of the persistence of this motile state, examining the influence of various parameters: the packing fraction of the medium (Section \ref{sub:MFPTpacking}), the magnitude of the active force (\ref{sub:MFPTforce}), and the maximum hinge angle (Section \ref{sub:MFPTangle}). Finally, in Section \ref{sec:discussion}, we discuss the implications of our findings and outline possible directions for future research.

\section{Active hinge model}
\label{sec:model}

\begin{figure}
\centering
  \includegraphics[width=\columnwidth]{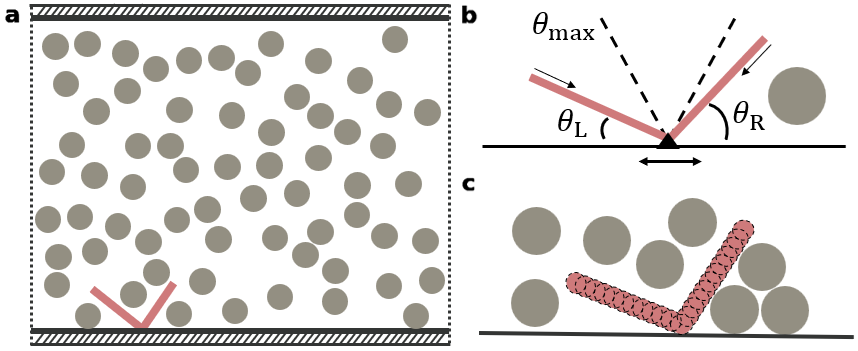}
  \caption{A schematic illustration of the active hinge model. (a) The active hinge, highlighted in red, is immersed in a bath of passive Brownian particles confined within a two-dimensional channel. (b) A detailed view of the active hinge.}
  \label{fgr:model}
\end{figure}

As a simplified description of the experimental system discussed in Son \textit{et al.}\cite{son2024dynamics}, we propose the \textit{active hinge model}, which focuses on the dynamics of the self-propelled particle (SPP) cluster after it is formed at the boundary of the system. The model features two identical self-propelled, overdamped rods connected by a sliding hinge and immersed in a bath of overdamped passive Brownian disks confined in a two-dimensional channel with periodic boundaries in the $x$ direction and rigid, confining walls in the $y$ direction. Each rod exerts a constant active force (self-propulsion) of magnitude $F_\mathrm{act}$ along its body toward the hinge, its orientation fluctuating as it collides with the passive disks. The hinge is attached to a channel wall, moving only along the $x$ direction. The whole setup is illustrated in Fig.~\ref{fgr:model}. Throughout this study, taking the diameter of the passive disk, $\sigma$, as the unit length, we fix the dimensions of the system at $L_x = L_y = 10\sigma$, and the length of each rod at $l = 2\sigma$.

In the SPP clusters observed in the experiment~\cite{son2024dynamics}, due to the circular curvature of the wall and the elliptical shape of the SPPs, the effective maximum angle formed by the wall and the body axis of an SPP at the cluster boundary decreased as the cluster size increased (see Fig.~\ref{fgr:angle}). To emulate this effect, we introduce an upper bound, $\theta_\mathrm{max}$, on the angle between each rod and the wall (so that $\theta_\mathrm{L,R} \le \theta_\mathrm{max}$), as indicated by dashed lines in Fig.~\ref{fgr:model}(b).

In accordance with these descriptions, the horizontal location $x_\mathrm{act}$, the angle $\theta$ formed by each rod and the wall, and the position $\mathbf{x}_\mathrm{pas}$ of each passive disk obey the overdamped Langevin equations
\begin{align}
    \mu^{-1}\dot{x}_{\mathrm{act}} &= F_{\mathrm{act}}\big[cos(\theta_\mathrm{L})-cos(\theta_\mathrm{R})\big] + \mathbf{F}_{\mathrm{int}}^{\mathrm{act}}\cdot\hat{\mathbf{x}},
    \label{eq:dynacttrans}
    \\
    \mu_{\mathrm{r}}^{-1}\dot{\theta} &= \Gamma_{\mathrm{int}},
    \label{eq:dynactrot}\\
    \mu^{-1}\dot{\mathbf{x}}_{\mathrm{pas}} &= \mathbf{F}_{\mathrm{int}}^{\mathrm{pas}} + \sqrt{2D}\mathbf{\xi}.
    \label{eq:dynpastrans}
\end{align}
Here $\mu$ is the mobility of the active hinge, $\mathbf{F}_{\mathrm{int}}^{\mathrm{act}}$ the total force exerted upon the active hinge by the passive disks, $\mu_\mathrm{r}$ the rotational mobility of each rod, $\Gamma_\mathrm{int}$ the total torque exerted upon each rod by the passive disks, $\mathbf{F}_{\mathrm{int}}^{\mathrm{pas}}$ the total force exerted upon a passive disk by the wall, the self-propelled rods, and the other passive disks, $D$ the diffusion coefficient, and $\xi$ the Gaussian white noise of unit amplitude. The values of $F_{\mathrm{int}}^{\mathrm{act}}$, $\Gamma_\mathrm{int}$, and $F_{\mathrm{int}}^{\mathrm{pas}}$ are calculated based on the following assumptions:
\begin{enumerate}
\item The self-propelled rods can be regarded as a rigid chain of identical disks, as illustrated in Fig.~\ref{fgr:model}(c). In our simulations, each rod is composed of $40$ disks of radius $0.1\sigma$.
\item A pair of disks separated by a center-to-center distance $r$ interact via the Weeks--Chandler--Andersen (WCA) potential
\begin{align}
U(r_{ij}) = \begin{cases}
4\epsilon\Big[\Big(\frac{\sigma_{ij}}{r_{ij}}\Big)^{12}-\Big(\frac{\sigma_{ij}}{r_{ij}}\Big)^6\Big] & \text{ if } r_{ij}\leq2^{1/6}\sigma_{ij}, \\ 
0 & \text{ otherwise. }
\end{cases}
\label{eq:wcapotential}
\end{align}
Here, $\sigma_{ij}$ represents the sum of the radii of two interacting objects. Also note that, in the case of disk--wall interactions, $r_{ij}$ represents the distance between the center of the disk and the wall. In our study, we use $\epsilon = 50$ and $\sigma_{ij} = \sigma$ for interactions between passive disks as well as disk--wall interactions, while $\epsilon = 5$ and $\sigma_{ij} = 0.6\sigma$ for interactions between passive disks and self-propelled rods. Note that these chosen values of $\varepsilon$ set the unit of force in our study.
\end{enumerate}

Once either $\theta_\mathrm{L}$ or $\theta_\mathrm{R}$ reaches zero, the corresponding rod stays attached to the wall since no passive disks can split them apart. Thus, the active hinge reaches an absorbing state once both angles become zero.

\section{Symmetry-breaking motility of the hinge}
\label{sec:symmetrybreaking}

\begin{figure}
\centering
  \includegraphics[width=\columnwidth]{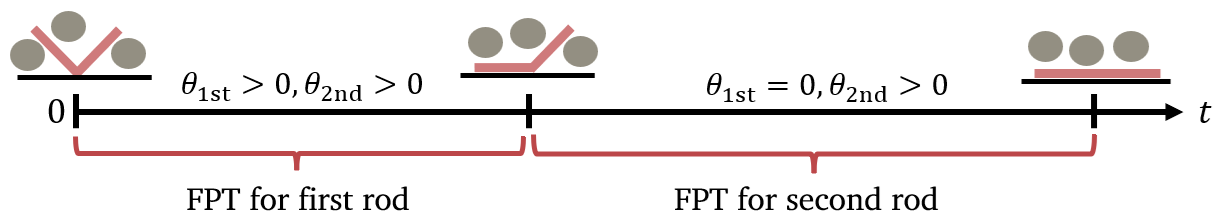}
  \caption{Design of the simulation. Starting from the symmetric hinge configuration, two first passage times (FPTs) can be defined depending on which rod is absorbed.}
  \label{fgr:fpt}
\end{figure}

We first demonstrate that the active hinge exhibits a motile state maintained by a symmetry-breaking mechanism. Toward this aim, we simulate the system starting from the symmetric configuration with $\theta_\mathrm{L,R} = \theta_\mathrm{max}$ (with the passive disks having reached a steady state under this constraint) and then measure the mean first passage times (MFPTs) for each angle to reach zero. Let us call the rod whose angle with the wall reaches zero first the `first rod', while the other is called the `second rod'. The MFPT of the first rod refers to the average period of time between the beginning of the observation and the absorption of the first rod. The MFPT of the second rod refers to the average period of time between the absorption of the first rod and that of the second rod. See Fig.~\ref{fgr:fpt} for a schematic of the idea and \hyperref[app:mfpt_surv]{Appendix} for how the MFPT is estimated.

\begin{figure}
\centering
\includegraphics[width=\columnwidth]{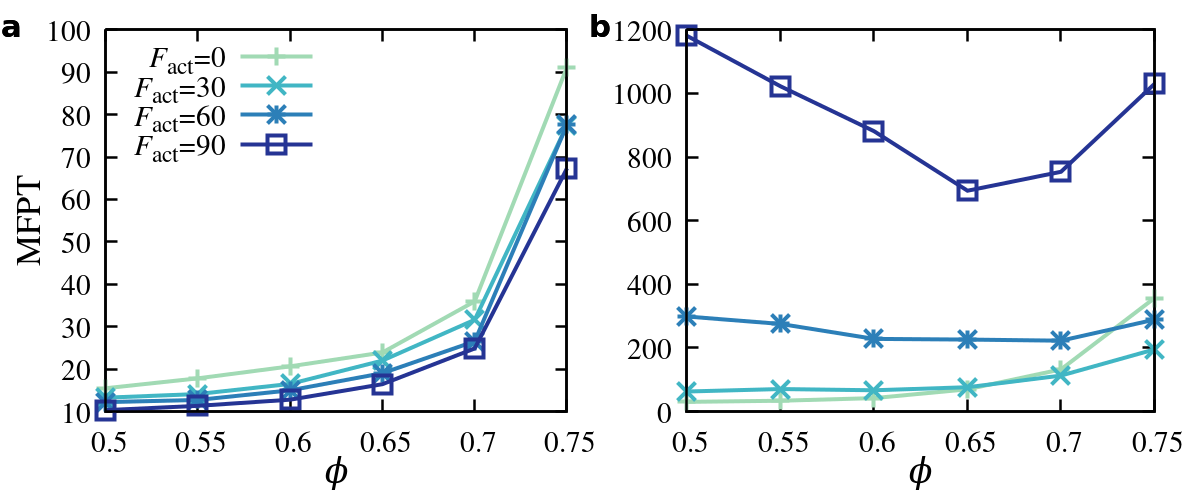}
\caption{MFPT of the (a) first and the (b) second rods as a function of the packing fraction $\phi$ for different self-propulsion strengths $F_\mathrm{act}$. We use $\theta_\mathrm{max}=60^\circ$. The lines are to guide the eye, and the error bars are smaller than the symbols.}
\label{fgr:MFPT_both}
\end{figure}

In Fig.~\ref{fgr:MFPT_both}, we observe that under various conditions the MFPT of the first rod is much shorter than that of the second rod. As shown in Fig.~\ref{fgr:MFPT_both}(a), the MFPT of the first rod monotonically increases with the packing fraction $\phi$ of the passive disks, reflecting that the increased collisions with the background medium slows down the absorption of the first rod. This is in contrast to the MFPT of the second rod, which exhibits nonmonotonic dependence on $\phi$ as shown in Fig.~\ref{fgr:MFPT_both}(b). This suggests multiple competing mechanisms that promote or inhibit the motile state of the hinge, which will be discussed in more detail in the next section.

\begin{figure}
\centering
\includegraphics[width=\columnwidth]{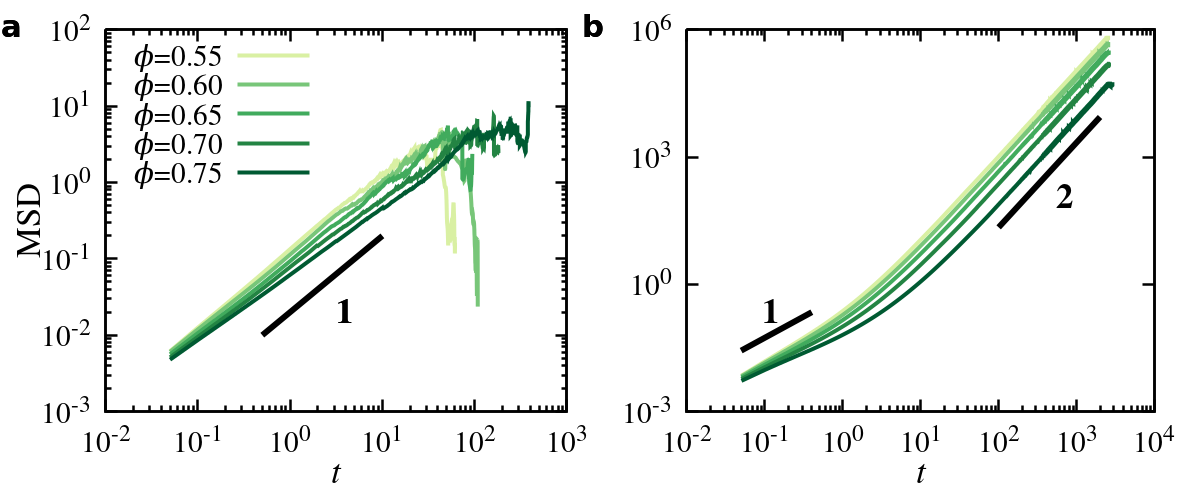}
\caption{MSD of the hinge for different packing fractions $\phi$ (a) up to the absorption of the first rod and (b) up to the absorption of the second rod. The thick black segments with numbers are for comparisons with normal diffusion ($\mathrm{MSD} \sim t$) and ballistic motion ($\mathrm{MSD} \sim t^2$). We use $F_\mathrm{act} = 90$ and $\theta_\mathrm{max} = 60^\circ$.}
\label{fgr:MSD}
\end{figure}

The mean squared displacement (MSD) of the hinge also corroborates the existence of the motile state. As Fig.~\ref{fgr:MSD}(a) shows, the hinge undergoes normal diffusion until the first rod is absorbed. Then, as shown in Fig.~\ref{fgr:MSD}(b), the motion stays ballistic for a much longer period of time, which corresponds to the persistent motile state. These results indicate that, once the first rod is absorbed, the hinge enters a motile state maintained by the broken symmetry, as was the case for the experimental system discussed in Son~\textit{et al.}~\cite{son2024dynamics}.

\section{Persistence of the motile state}
\label{sec:results}

To further investigate the multiple competing mechanisms that contribute to the persistence of the motile state, from now on we focus on the case where the first rod is absorbed from the beginning while the initial angle of the second rod is given by $\theta_\mathrm{max}$. Under this condition, we examine the MFPT of the second rod as an indicator of the lifetime of the motile state for various values of the packing fraction $\phi$, the active force magnitude $F_\mathrm{act}$, and the maximum angle $\theta_\mathrm{max}$.

\subsection{Effects of the packing fraction}
\label{sub:MFPTpacking}

\begin{figure}
\centering
\includegraphics[width=\columnwidth]{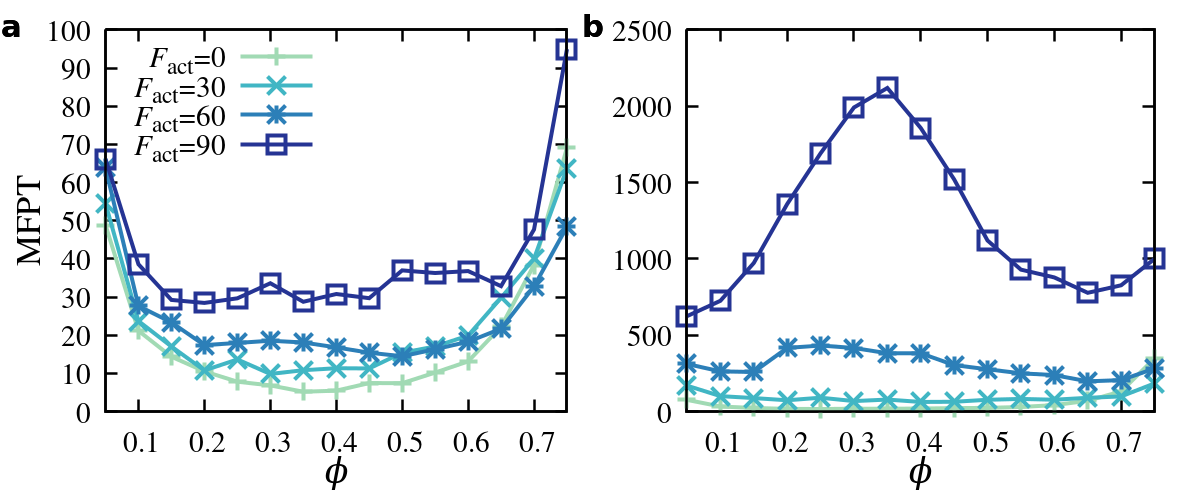}
\caption{Lifetime of the motile state indicated by the MFPT as a function of the packing fraction for different values of the active force magnitude $F_\mathrm{act}$ and the maximim angle $\theta_\mathrm{max}$. (a) For $\theta_\mathrm{max}=45^\circ$, the MFPT only exhibits U-shaped curves. (b) For $\theta_\mathrm{max}=60^\circ$, the MFPT shows a peak in the intermediate range of $\phi$ when $F_\mathrm{act}$ is sufficiently large. The lines are to guide the eye, and the error bars are smaller than the symbols.}
\label{fgr:MFPT_packing}
\end{figure}

The packing fraction $\phi$, defined as the total area fraction covered by the passive disks, controls how frequently the active hinge interacts with the granular medium. In Fig.~\ref{fgr:MFPT_packing}, we show the $\phi$ dependence of the lifetime of the motile state for different values of $F_\mathrm{act}$ and $\theta_\mathrm{max}$. As shown in Fig.~\ref{fgr:MFPT_packing}(a) for $\theta_\mathrm{max}=45^\circ$, when $\theta_\mathrm{max}$ is small, the MFPT exhibits a U-shaped behavior as a function of $\phi$. On the other hand, as shown in Fig.~\ref{fgr:MFPT_packing}(b) for $\theta_\mathrm{max}=60^\circ$, we observe a peak of the MFPT in the intermediate range of $\phi$ when both $\theta_\mathrm{max}$ and $F_\mathrm{act}$ are large enough.

\begin{figure}
\centering
\includegraphics[width=\columnwidth]{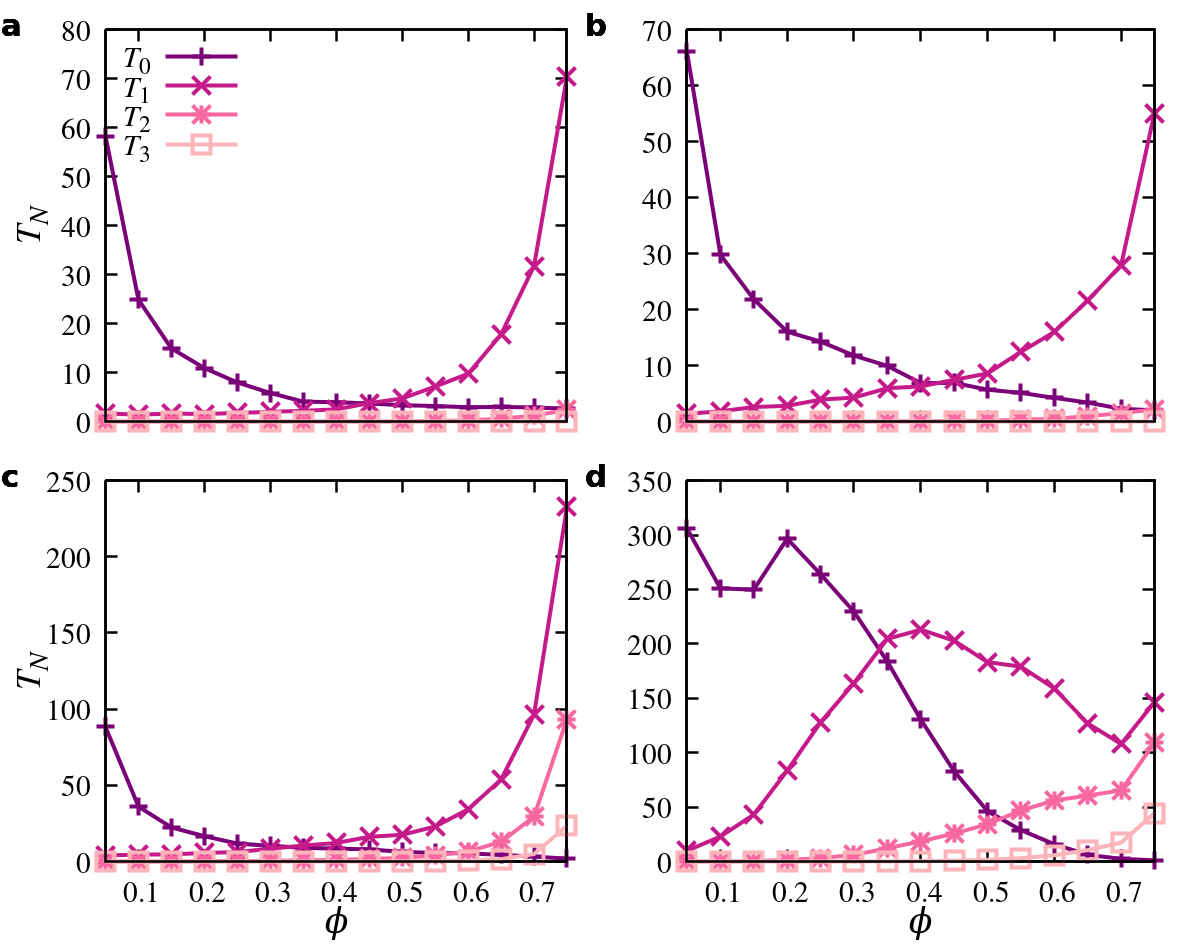}
\caption{Average time spent in each coarse-grained state $N$ until absorption as the packing fraction is varied. We show the results for (a) $\theta_\mathrm{max}=45^\circ$ and $F_\mathrm{act}=0$, (b) $\theta_\mathrm{max}=45^\circ$ and $F_\mathrm{act}=60$, (c) $\theta_\mathrm{max}=60^\circ$ and $F_\mathrm{act}=0$, and (d) $\theta_\mathrm{max}=60^\circ$ and $F_\mathrm{act}=60$. The lines are to guide the eye, and the error bars are smaller than the symbols.}
\label{fgr:MFPT_terms_packing}
\end{figure}

To further investigate the origins of these behaviors, we represent the state of the active hinge using the number of passive disks whose center falls in the circular sector of radius $l$ (\textit{i.e.}, the length of each rod) between the surviving rod and the channel wall. Denoting this number by $N$, we estimate the average time spent in each state $T_N$ from the simulation data using the method described in \hyperref[app:mfpt_surv]{Appendix}. This allows us to examine the detailed structure of the MFPT, which is equal to the sum of $T_N$ over all possible values of $N$.

As shown in the upper pannel of Fig.~\ref{fgr:MFPT_terms_packing}, when $\theta_\mathrm{max} = 45^\circ$, only $T_0$ and $T_1$ account for most of the MFPT. This is because $\theta_\mathrm{max}$ is not large enough to allow more than a single passive disk between the rod and the wall. The monotonic decrease of $T_0$ with increasing $\phi$ can be attributed to the increasing rate of collision with the passive disks that accelerate the rotational dynamics of the rod. Meanwhile, the monotonic increase of $T_1$ with increasing $\phi$ is due to the crowding of the channel slowing down the escape of the passive disk between the rod and the wall. These behaviors are qualitatively the same regardless of whether the hinge is passive [$F_\mathrm{act} = 0$, as in Fig.~\ref{fgr:MFPT_terms_packing}(a)] or active [$F_\mathrm{act} = 60$, as in Fig.~\ref{fgr:MFPT_terms_packing}(b)]. Even if we increase $\theta_\mathrm{max}$ to $60^\circ$, we still observe essentially similar behaviors as long as $F_\mathrm{act}$ is weak, except that $T_3$ and $T_4$ make nonnegligible contributions to the MFPT as more passive disks are allowed between the rod and the wall.

Notable changes occur only when both $\theta_\mathrm{max}$ and $F_\mathrm{act}$ are large enough, as shown in Fig.~\ref{fgr:MFPT_terms_packing}(d) for $\theta_\mathrm{max} = 60^\circ$ and $F_\mathrm{act} = 60$. In this case, the configuration with $N = 1$ achieves a prolonged lifetime even for small values of $\phi$, with $T_1$ dominating the MFPT for an intermediate range of $\phi$. This can be attributed to the positive feedback between cluster asymmetry and motility analogous to the mechanism that induced the symmetry-breaking motility in the experiment of Son \textit{et al.}\cite{son2024dynamics}, which was illustrated in Fig.~\ref{fgr:angle}. The mechanism requires that the cluster moves quickly enough to keep the passive disk stuck between the rod and the wall, which is more likely when the cluster has a strong propulsion (via large $F_\mathrm{act}$) and is allowed to be highly asymmetric (via large $\theta_\mathrm{max}$). However, as the system becomes more crowded, stronger pressure is applied on the rod by the disks above the hinge, which tends to decrease the rod--wall angle, weaken the asymmetry of the motile cluster, and slow it down. This makes it easier for the passive disk trapped between the rod and the wall to escape, and as soon as $N$ changes from $1$ to $0$, the pressure quickly drives the surviving rod to absorption. This explains why both $T_0$ and $T_1$ decrease in the upper part of the intermediate range of $\phi$, completing a peak of the MFPT.

\begin{figure}
\centering
\includegraphics[width=\columnwidth]{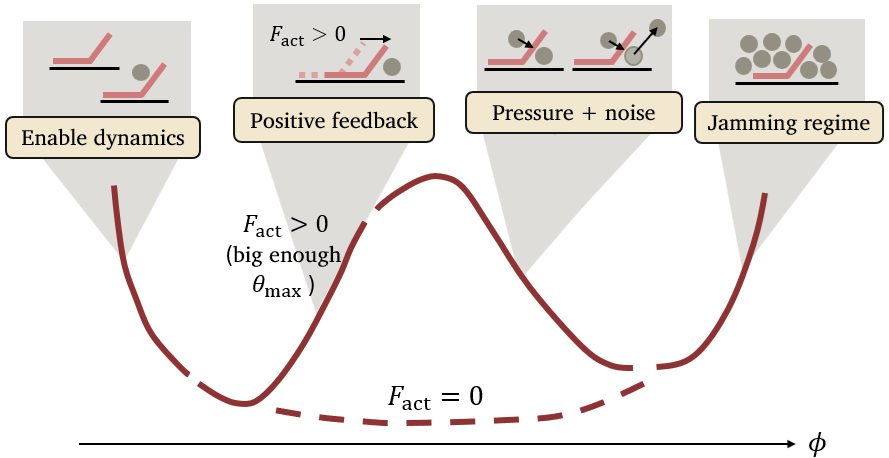}
\caption{Schematic of the four mechanisms contributing to the persistence of the motile state as $\phi$ is varied. The dark red lines represent the behaviors of the MFPT of the active hinge. When $\phi$ is very small, introduction of passive particles enables rotational diffusion, lowering the MFPT. As $\phi$ increases, it is more likely for a single passive disk to be trapped between the surviving rod and the wall, inducing persistent motion via positive feedback between cluster asymmetry and motility. If $\phi$ is increased further, pressure applied by the disks above the hinge slows its motion and reduces the MFPT. In the jamming regime ($\phi \gtrsim 0.7$), the whole dynamics slow down, causing the MFPT to rise again. The peak of the MFPT in the intermediate regime is not observed when $\theta_\mathrm{max}$ or $F_\mathrm{act}$ is not large enough.}
\label{fgr:pf_explanation}
\end{figure}

Based on these observations, we can identify four competing mechanisms that contribute to the lifetime of the motile state: (i) rotational diffusion of the surviving rod, whose activation causes a decrease in MFPT with $\phi$ when $\phi$ is small; (ii) positive feedback between cluster motility and asymmetry, which causes an increase in MFPT with $\phi$ when $\phi$ is in the lower intermediate range; (iii) pressure by the disks above the hinge combined with the noisy escape of the trapped passive disk, which causes a decrease in MFPT with $\phi$ when $\phi$ is in the upper intermediate range; (iv) jamming that slows down the dynamics and increases the MFPT when $\phi$ is very high. The second and third of these mechanisms become significant only when both $F_\mathrm{act}$ and $\theta_\mathrm{max}$ are large enough, allowing strong cluster asymmetry and motility to develop. A schematic illustration of these mechanisms is provided in Fig.~\ref{fgr:pf_explanation}.

\subsection{Effects of the active force magnitude}
\label{sub:MFPTforce}

As already shown in Fig.~\ref{fgr:MFPT_packing}, for most values of $\phi$, the MFPT tends to increase monotonically with $F_\mathrm{act}$. This may be because, when $F_\mathrm{act}$ is greater, the active hinge in the $N = 1$ state moves faster, and it takes longer for the trapped passive disk to escape.

\begin{figure}
\centering
\includegraphics[width=9cm]{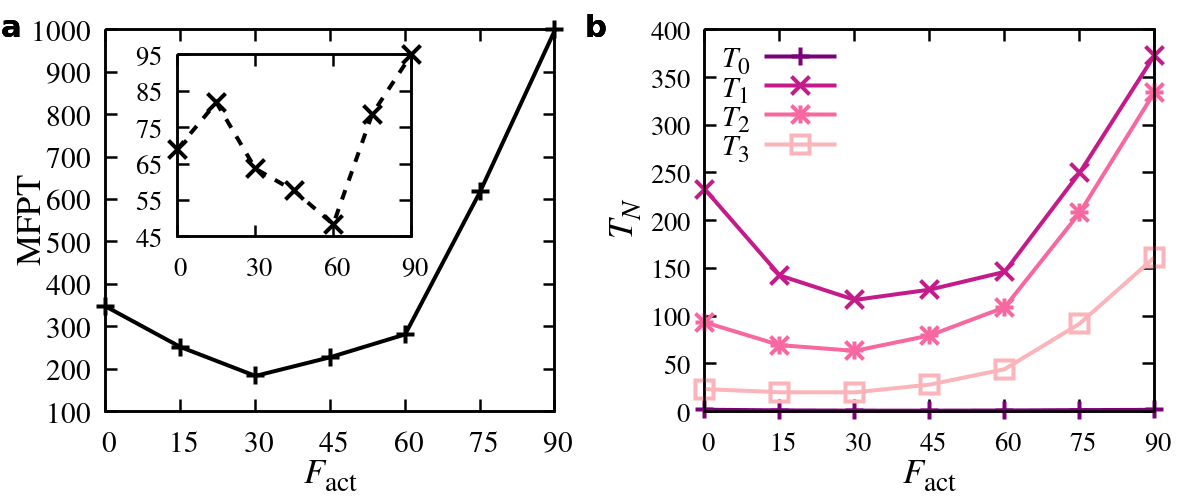}
\caption{(a) MFPT as a function of the active force magnitude at $\phi = 0.75$ and $\theta_\mathrm{max}=60^\circ$ (the inset is for $\theta_\mathrm{max}=45^\circ$). The motile state achieves the longest lifetime for intermediate strengths of the active force. (b) Average time spent at each coarse-grained state $N$ until absorption as a function of the active force magnitude at $\phi = 0.75$ and $\theta_\mathrm{max}=60^\circ$. The lines are to guide the eye, and the error bars are smaller than the symbols.}
\label{fgr:MFPT_force}
\end{figure}

However, when $\phi$ is very high ($\phi \gtrsim 0.7$) and $\theta_\mathrm{max} = 60^\circ$, one observes that the MFPT does not show a monotonic growth as we increase $F_\mathrm{act}$. This nonmonotonic behavior is examined in more detail in Fig.~\ref{fgr:MFPT_force}(a). While the MFPT does not have a clear dependence on $F_\mathrm{act}$ for $\theta_\mathrm{max} = 45^\circ$ (inset), it is clearly minimized at an intermediate value of $F_\mathrm{act}$ when $\theta_\mathrm{max} = 60^\circ$. To examine this phenomenon in more detail, we show the behaviors of different $T_N$'s as functions of $F_\mathrm{act}$ in Fig.~\ref{fgr:MFPT_force}(b). We observe that the $N = 0$ state has only negligible contributions to the MFPT, which implies that the active hinge is absorbed as soon as the passive disk trapped between the rod and the wall escapes. The $N = 1$ state dominates the MFPT for small $F_\mathrm{act}$, playing the role of the bottleneck that prevents the active hinge from absorption. When $F_\mathrm{act}$ increases, the active hinge spends considerably less time in the $N = 1$ state, indicating that it becomes much easier for the trapped disk to escape. Then, when $F_\mathrm{act}$ increases even further, the lifetime of every state with $N \ge 1$ grows rapidly, reflecting the strong trapping effect in the regime.

\begin{figure}
\centering
\includegraphics[width=4.5cm]{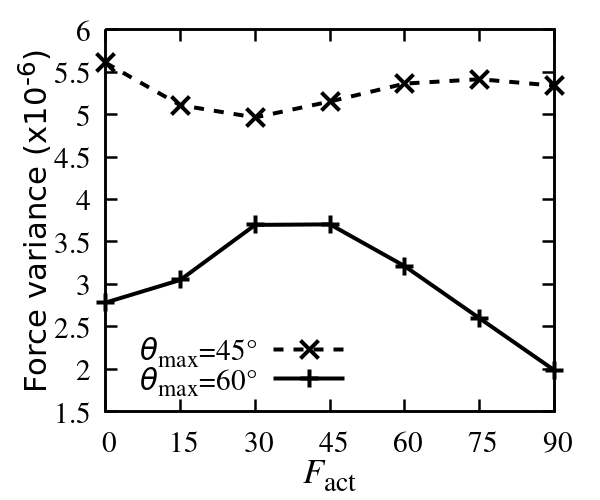}
\caption{Variance of the total $x$-directional force acting on the active hinge in the $N = 1$ state at $\phi=0.75$. For $\theta_\mathrm{max} = 60^\circ$, the variance peaks at intermediate values of $F_\mathrm{act}$, indicating stronger noise affecting the trapped disk in this regime. The lines are to guide the eye, and the error bars are smaller than the symbols.}
\label{fgr:variance_force}
\end{figure}

Why is it easier to overcome the bottleneck in the $N = 1$ state when $F_\mathrm{act}$ is in the intermediate range? In Fig.~\ref{fgr:variance_force}, we show the variance of the $x$-directional total force (including both the self-propulsion forces and the rod--disk interactions) applied on the active hinge in the $N = 1$ state. While the variance does not change that much for $\theta_\mathrm{max} = 45^\circ$, it exhibits a pronounced peak when $\theta_\mathrm{max} = 60^\circ$. This can be attributed to a tradeoff between the fluctuation enhancement by the stronger self-propulsion and the fluctuation inhibition by the jamming effects as $F_\mathrm{act}$ increases. Since the force variance would be proportional to the effective magnitude of the noise that affects the trapped passive disk, its escape time would be minimized for the intermediate range of $F_\mathrm{act}$, leading to the nonmonotonic behaviors of the MFPT.

\subsection{Effects of the maximum angle}
\label{sub:MFPTangle}

Lastly, we discuss how $\theta_\mathrm{max}$ affects the lifetime of the motile state. As noted in Sec.~\ref{sub:MFPTpacking}, higher $\theta_\mathrm{max}$ makes it easier for the system to form a symmetry-breaking motile structure, so the MFPT would naturally increase with $\theta_\mathrm{max}$. Here, we focus on the case $\phi = 0.75$, which corresponds to the packing fraction used in the experiment of Son \textit{et al.}\cite{son2024dynamics}, although the results remain qualitatively similar even for $\phi$ as low as $0.3$.

\begin{figure}
\centering
\includegraphics[width=\columnwidth]{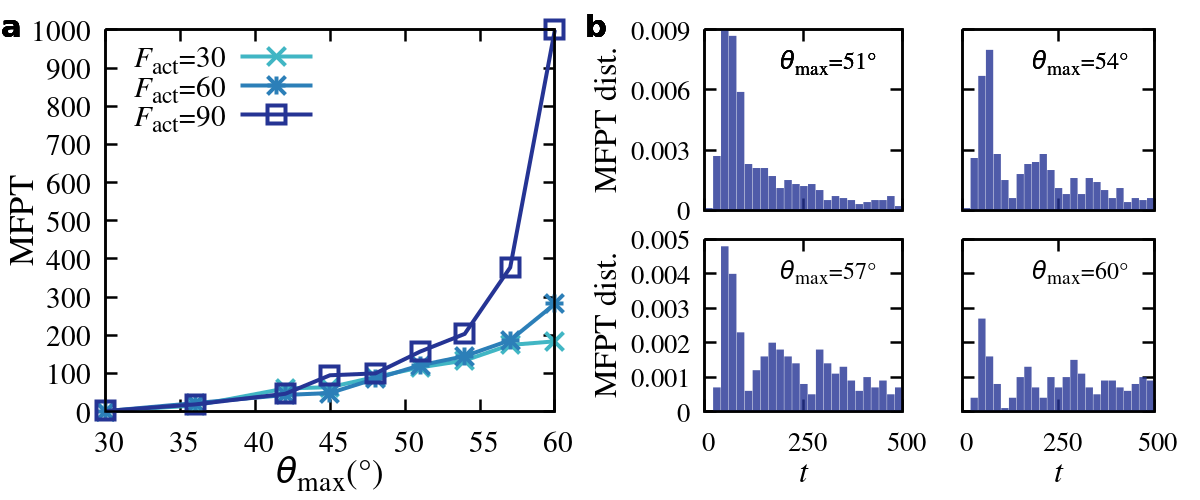}
\caption{(a) MFPT as a function of the maximum angle at $\phi=0.75$ and for different active force magnitudes. For $F_\mathrm{act}=90$, the lifetime of the motile state increases rapidly when $\theta_\mathrm{max}$ is increased beyond $\theta_\mathrm{max} \approx 53^\circ$, so that three passive disks can form a stable triangular configuration between the rod and the wall. The lines are to guide the eye, and the error bars are smaller than the symbols. (b) MFPT distributions for various values of $\theta_\mathrm{max}$ at $\phi=0.75$ and $F_\mathrm{act}=90$. Multiple peaks are observable for $\theta_\mathrm{max} \gtrsim 53^\circ$.}
\label{fgr:MFPT_ang}
\end{figure}

In Fig.~\ref{fgr:MFPT_ang}(a), we show the MFPT as a function of $\theta_\mathrm{max}$ for three different values of the active force magnitude. While the MFPT generically increases with $\theta_\mathrm{max}$, it exhibits a rapid growth for $F_\mathrm{act} = 90$ in the regime $\theta_\mathrm{max}\gtrsim 53^\circ$. As shown in Fig.~\ref{fgr:MFPT_ang}(b), the MFPT distribution develops multiple peaks in the same regime, which reflects the changed configuration of the passive particle cluster that prevents the active hinge from reaching the absorbing state.

\begin{figure}
\centering
\includegraphics[width=\columnwidth]{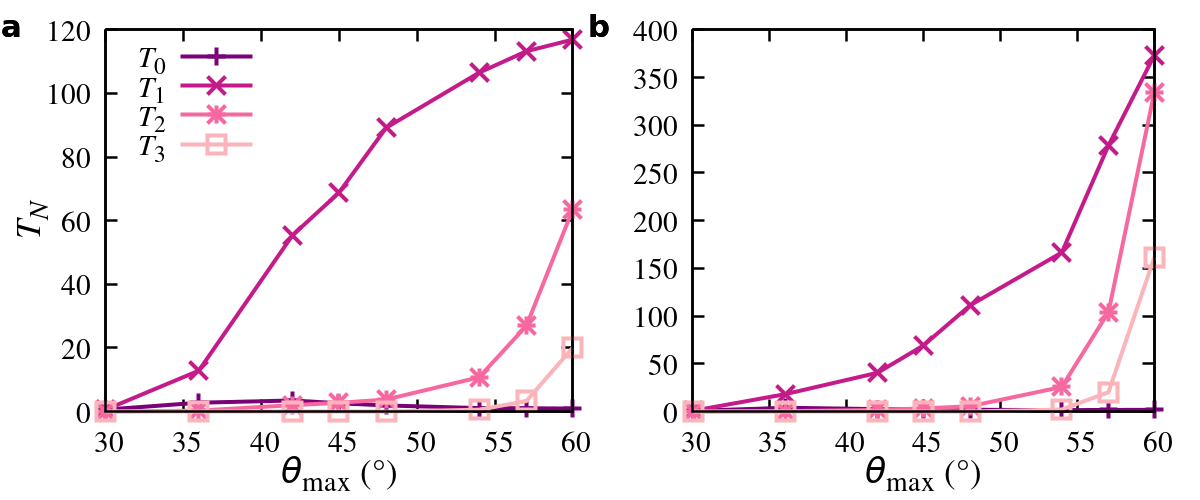}
\caption{Average time spent in each coarse-grained state $N$ until absorption as a function of the maximum angle when $\phi=0.75$ and (a) $F_\mathrm{act}=30$ or (b) $F_\mathrm{act}=90$. The lines are to guide the eye, and the error bars are smaller than the symbols.}
\label{fgr:MFPT_terms_ang}
\end{figure}

To gain more intuition into the origins of these behaviors, we examine the average time spent in each coarse-grained state $N$ until absortion in Fig.~\ref{fgr:MFPT_terms_ang}. As shown by both subfigures, if $\theta_\mathrm{max} \lesssim 53^\circ$, the $N = 1$ state dominates the MFPT. This is because the circular sector between the rod and the wall is too narrow to allow any stable cluster of multiple passive disks. If $\theta_\mathrm{max} \gtrsim 53^\circ$ and $F_\mathrm{act} = 30$, then as shown in Fig.~\ref{fgr:MFPT_terms_ang}(a), the $N = 1$ state still plays the role of a bottleneck while contributions from the other states remain small. However, if $F_\mathrm{act} = 90$, Fig.~\ref{fgr:MFPT_terms_ang}(b) shows that the system spends a prolonged period of time in the $N = 2$ and $N = 3$ states. This is because three passive disks can form an extremely stable triangular cluster in the regime illustrated in Fig.~\ref{fgr:angle}(b). Even if a single disk happens to leave the cluster, another disk will join the cluster with high probability. Moreover, the triangular cluster can also contribute to the $N = 2$ state since, even without the cluster disintegration, one of its disks can still fall outside the circular sector used to determine the value of $N$.

If $F_\mathrm{act}$ is not large enough, the forces trapping the cluster cannot hold it together for long, as reflected in the little time spent in the $N = 3$ state in Fig.~\ref{fgr:MFPT_terms_ang}(a) for $F_\mathrm{act} = 30$.

As discussed in Sec.~\ref{sec:model} (and illustrated in Fig.~\ref{fgr:angle}), $\theta_\mathrm{max}$ emulates the effects of the number of SPPs forming the cluster in the experiment of Son~\textit{et al.}\cite{son2024dynamics}. Our results corroborate the study's observation that the lifetime of the motile state exhibits an abrupt change as the number of SPPs is varied.

\section{Conclusions}
\label{sec:discussion}

Using extensive simulations of the active hinge model, which simplifies the dynamics of the self-propelled particle cluster at the boundary of a vibrated granular medium, we investigated how the packing fraction of the medium, the active force of the rods, and the maximum angle between each rod and the boundary affect the motion of the cluster. Our results show that, even if the initial configuration of the hinge is fully symmetric, it takes much longer for the second rod to be absorbed compared to the first one due to the positive feedback between the asymmetry of the hinge and its motility. When the active force is strong enough, persistence of this mechanism exhibits a double-dipped dependence on the packing fraction, demonstrating the interplay of the positive feedback mechanism and the pressure exerted by the granular particles. Moreover, in the regime of sufficiently high packing fraction and maximum angle, persistence of motion is minimized at an intermediate value of the active force, which reflects the tradeoff between self-propulsion and jamming effects. Finally, as the maximum angle grows, we observe that the lifetime of the motile state increases by orders of magnitude since the configuration of the passive granular medium sustaining the motile state changes abruptly. This is analogous to how the active boundary cluster in the granular medium confined within a circular arena suddenly loses its motility as the number of active particles increases~\cite{son2024dynamics}.

These results demonstrate how an artificial system composed of a mixture of active and passive particles can self-organize to exhibit various dynamical modes by changing the number and physical properties of particles. In particular, if the boundary of the system is reshaped into a movable circular arena instead of the straight channel considered here, the motile and the immotile states of the active hinge will lead to straight and chiral motion of the whole superstructure, respectively. As suggested by our results, the switching between the two modes can be done by changing the area of the arena (which in turn changes the packing fraction) or the active force. Designing a mobile and steerable superstructure based on this principle may provide an interesting alternative to those utilizing interactions between active particles and soft membranes~\cite{deblais2018boundaries, boudet2021collections,lee2023complex}.

\section*{Appendix: MFPT estimation}
\label{app:mfpt_surv}

The most straightforward way to estimate the MFPT is to obtain the sample mean of the first-passage time via numerous runs of the simulation. However, when the motile state exhibits a prolonged lifetime, there may be too few samples that reach the absorbing state, making the sample mean unreliable. To address this issue, we infer the MFPT from the survival probability of the rod, which is defined as
\begin{equation}
    \mathrm{SP}(t) = \frac{1}{N_\mathrm{sim}}\sum_{i=0}^{N_\mathrm{sim}}H(\tau_i-t).
    \label{eq:survival_prob}
\end{equation}
Here, $N_\mathrm{sim}$ is the number of simulation runs, $\tau_i$ is the absorbing time of the $i$th sample, and $H(x)$ is the Heaviside step function, which satisfies $H(\tau_i-t)=1$ only before absorption of the rod ($t < \tau_i$) and $H(\tau_i-t)=0$ otherwise. Assuming that the transient effects of the initial state quickly die out, we can expect the rod absorption to be a Poisson point process, so that the survival probability exhibits an exponential decay
\begin{equation}
    \mathrm{SP}(t) \propto \mathrm{e}^{-\gamma t}.
    \label{eq:survival_exp}
\end{equation}
After fitting the simulation data to infer the value $\gamma$, we use $\gamma^{-1}$ as the estimator of the MFPT $\langle \tau \rangle$.

The average time $T_N$ spent in a coarse-grained state $N$, where $N$ is the number of passive disks below the rod, can be estimated by a similar method. For each value of $N$, we define
\begin{equation}
    S(N,t) = \sum_{i=0}^{N_\mathrm{sim}}H\!\left(\sum_{k}\delta_{n_i(k\Delta t),N}\Delta t-t\right),
    \label{eq:survival_prob_N}
\end{equation}
where $\Delta t$ is the length of each discretized time step, and $n_i(k\Delta t)$ is the coarse-grained state of the $i$th sample at the $k$th time step. Then we infer $T_N$ by fitting the data to $S(N,t) \sim \mathrm{e}^{-t/T_N}$.

\section*{Author contributions}
\textbf{L. G. Rigon:} Conceptualization, Software, Validation, Formal analysis, Investigation, Writing -- Original draft, Visualization.\\
\textbf{Y. Baek:} Conceptualization, Methodology, Supervision, Writing -- Review \& editing.

\section*{Conflicts of interest}
There are no conflicts to declare.

\section*{Data availability}
The computer codes used to generate the results of this paper are available at \url{https://github.com/lgrigon/ActiveHingeModel}.

\section*{Acknowledgements}
This work was supported by the National Research Foundation of Korea (NRF) grants (RS-2021-017476 and RS-2023-00278985) funded by the Ministry of Science and ICT (MSIT) of the Korea government.



\balance


\bibliography{rsc} 

\providecommand*{\mcitethebibliography}{\thebibliography}
\csname @ifundefined\endcsname{endmcitethebibliography}
{\let\endmcitethebibliography\endthebibliography}{}
\begin{mcitethebibliography}{46}
\providecommand*{\natexlab}[1]{#1}
\providecommand*{\mciteSetBstSublistMode}[1]{}
\providecommand*{\mciteSetBstMaxWidthForm}[2]{}
\providecommand*{\mciteBstWouldAddEndPuncttrue}
  {\def\EndOfBibitem{\unskip.}}
\providecommand*{\mciteBstWouldAddEndPunctfalse}
  {\let\EndOfBibitem\relax}
\providecommand*{\mciteSetBstMidEndSepPunct}[3]{}
\providecommand*{\mciteSetBstSublistLabelBeginEnd}[3]{}
\providecommand*{\EndOfBibitem}{}
\mciteSetBstSublistMode{f}
\mciteSetBstMaxWidthForm{subitem}
{(\emph{\alph{mcitesubitemcount}})}
\mciteSetBstSublistLabelBeginEnd{\mcitemaxwidthsubitemform\space}
{\relax}{\relax}

\bibitem[Ramaswamy(2010)]{RamaswamyARCMP2010}
S.~Ramaswamy, \emph{Annu. Rev. Condens. Matter Phys.}, 2010, \textbf{1}, 323 -- 345\relax
\mciteBstWouldAddEndPuncttrue
\mciteSetBstMidEndSepPunct{\mcitedefaultmidpunct}
{\mcitedefaultendpunct}{\mcitedefaultseppunct}\relax
\EndOfBibitem
\bibitem[Marchetti \emph{et~al.}(2013)Marchetti, Joanny, Ramaswamy, Liverpool, Prost, Rao, and Simha]{MarchettiRMP2013}
M.~C. Marchetti, J.~F. Joanny, S.~Ramaswamy, T.~B. Liverpool, J.~Prost, M.~Rao and R.~A. Simha, \emph{Rev. Mod. Phys.}, 2013, \textbf{85}, 1143--1189\relax
\mciteBstWouldAddEndPuncttrue
\mciteSetBstMidEndSepPunct{\mcitedefaultmidpunct}
{\mcitedefaultendpunct}{\mcitedefaultseppunct}\relax
\EndOfBibitem
\bibitem[Bechinger \emph{et~al.}(2016)Bechinger, Di~Leonardo, L\"owen, Reichhardt, Volpe, and Volpe]{BechingerRMP2016}
C.~Bechinger, R.~Di~Leonardo, H.~L\"owen, C.~Reichhardt, G.~Volpe and G.~Volpe, \emph{Rev. Mod. Phys.}, 2016, \textbf{88}, 045006\relax
\mciteBstWouldAddEndPuncttrue
\mciteSetBstMidEndSepPunct{\mcitedefaultmidpunct}
{\mcitedefaultendpunct}{\mcitedefaultseppunct}\relax
\EndOfBibitem
\bibitem[Ramaswamy(2017)]{RamaswamyJSM2017}
S.~Ramaswamy, \emph{J. Stat. Mech.: Theor. Exp.}, 2017, \textbf{2017}, 054002\relax
\mciteBstWouldAddEndPuncttrue
\mciteSetBstMidEndSepPunct{\mcitedefaultmidpunct}
{\mcitedefaultendpunct}{\mcitedefaultseppunct}\relax
\EndOfBibitem
\bibitem[J\"{u}licher \emph{et~al.}(2018)J\"{u}licher, Grill, and Salbreux]{JulicherRPP2018}
F.~J\"{u}licher, S.~W. Grill and G.~Salbreux, \emph{Rep. Prog. Phys.}, 2018, \textbf{81}, 076601\relax
\mciteBstWouldAddEndPuncttrue
\mciteSetBstMidEndSepPunct{\mcitedefaultmidpunct}
{\mcitedefaultendpunct}{\mcitedefaultseppunct}\relax
\EndOfBibitem
\bibitem[Gompper \emph{et~al.}(2020)Gompper, Winkler, and {Speck {\em et al.}}]{GompperJPCM2020}
G.~Gompper, R.~G. Winkler and T.~{Speck {\em et al.}}, \emph{J. Phys.: Conden. Matter}, 2020, \textbf{32}, 193001\relax
\mciteBstWouldAddEndPuncttrue
\mciteSetBstMidEndSepPunct{\mcitedefaultmidpunct}
{\mcitedefaultendpunct}{\mcitedefaultseppunct}\relax
\EndOfBibitem
\bibitem[Bowick \emph{et~al.}(2022)Bowick, Fakhri, Marchetti, and Ramaswamy]{BowickPRX2022}
M.~J. Bowick, N.~Fakhri, M.~C. Marchetti and S.~Ramaswamy, \emph{Phys. Rev. X}, 2022, \textbf{12}, 010501\relax
\mciteBstWouldAddEndPuncttrue
\mciteSetBstMidEndSepPunct{\mcitedefaultmidpunct}
{\mcitedefaultendpunct}{\mcitedefaultseppunct}\relax
\EndOfBibitem
\bibitem[te~Vrugt and Wittkowski(2025)]{VrugtEPJE2025}
M.~te~Vrugt and R.~Wittkowski, \emph{Eur. Phys. J. E}, 2025, \textbf{48}, 12\relax
\mciteBstWouldAddEndPuncttrue
\mciteSetBstMidEndSepPunct{\mcitedefaultmidpunct}
{\mcitedefaultendpunct}{\mcitedefaultseppunct}\relax
\EndOfBibitem
\bibitem[Angelani and Leonardo(2010)]{AngelaniNJP2010}
L.~Angelani and R.~D. Leonardo, \emph{New J. Phys.}, 2010, \textbf{12}, 113017\relax
\mciteBstWouldAddEndPuncttrue
\mciteSetBstMidEndSepPunct{\mcitedefaultmidpunct}
{\mcitedefaultendpunct}{\mcitedefaultseppunct}\relax
\EndOfBibitem
\bibitem[Kaiser \emph{et~al.}(2014)Kaiser, Peshkov, Sokolov, ten Hagen, L\"owen, and Aranson]{KaiserPRL2014}
A.~Kaiser, A.~Peshkov, A.~Sokolov, B.~ten Hagen, H.~L\"owen and I.~S. Aranson, \emph{Phys. Rev. Lett.}, 2014, \textbf{112}, 158101\relax
\mciteBstWouldAddEndPuncttrue
\mciteSetBstMidEndSepPunct{\mcitedefaultmidpunct}
{\mcitedefaultendpunct}{\mcitedefaultseppunct}\relax
\EndOfBibitem
\bibitem[Mallory \emph{et~al.}(2014)Mallory, Valeriani, and Cacciuto]{MalloryPRE2014}
S.~A. Mallory, C.~Valeriani and A.~Cacciuto, \emph{Phys. Rev. E}, 2014, \textbf{90}, 032309\relax
\mciteBstWouldAddEndPuncttrue
\mciteSetBstMidEndSepPunct{\mcitedefaultmidpunct}
{\mcitedefaultendpunct}{\mcitedefaultseppunct}\relax
\EndOfBibitem
\bibitem[Smallenburg and L\"owen(2015)]{SmallenburgPRE2015}
F.~Smallenburg and H.~L\"owen, \emph{Phys. Rev. E}, 2015, \textbf{92}, 032304\relax
\mciteBstWouldAddEndPuncttrue
\mciteSetBstMidEndSepPunct{\mcitedefaultmidpunct}
{\mcitedefaultendpunct}{\mcitedefaultseppunct}\relax
\EndOfBibitem
\bibitem[Koumakis \emph{et~al.}(2013)Koumakis, Lepore, Maggi, and {Di Leonardo}]{KoumakisNComms2013}
N.~Koumakis, A.~Lepore, C.~Maggi and R.~{Di Leonardo}, \emph{Nat. Commun.}, 2013, \textbf{4}, 2588\relax
\mciteBstWouldAddEndPuncttrue
\mciteSetBstMidEndSepPunct{\mcitedefaultmidpunct}
{\mcitedefaultendpunct}{\mcitedefaultseppunct}\relax
\EndOfBibitem
\bibitem[Angelani \emph{et~al.}(2009)Angelani, Di~Leonardo, and Ruocco]{AngelaniPRL2009}
L.~Angelani, R.~Di~Leonardo and G.~Ruocco, \emph{Phys. Rev. Lett.}, 2009, \textbf{102}, 048104\relax
\mciteBstWouldAddEndPuncttrue
\mciteSetBstMidEndSepPunct{\mcitedefaultmidpunct}
{\mcitedefaultendpunct}{\mcitedefaultseppunct}\relax
\EndOfBibitem
\bibitem[Sokolov \emph{et~al.}(2010)Sokolov, Apodaca, Grzybowski, and Aranson]{SokolovPNAS2010}
A.~Sokolov, M.~M. Apodaca, B.~A. Grzybowski and I.~S. Aranson, \emph{Proc. Natl. Acad. Sci. USA}, 2010, \textbf{107}, 969--974\relax
\mciteBstWouldAddEndPuncttrue
\mciteSetBstMidEndSepPunct{\mcitedefaultmidpunct}
{\mcitedefaultendpunct}{\mcitedefaultseppunct}\relax
\EndOfBibitem
\bibitem[{Di Leonardo} \emph{et~al.}(2010){Di Leonardo}, Angelani, Dell'Arciprete, Ruocco, Iebba, Schippa, Conte, Mecarini, Angelis, and Fabrizio]{DiLeonardoPNAS2010}
R.~{Di Leonardo}, L.~Angelani, D.~Dell'Arciprete, G.~Ruocco, V.~Iebba, S.~Schippa, M.~P. Conte, F.~Mecarini, F.~D. Angelis and E.~D. Fabrizio, \emph{Proc. Natl. Acad. Sci. USA}, 2010, \textbf{107}, 9541--9545\relax
\mciteBstWouldAddEndPuncttrue
\mciteSetBstMidEndSepPunct{\mcitedefaultmidpunct}
{\mcitedefaultendpunct}{\mcitedefaultseppunct}\relax
\EndOfBibitem
\bibitem[Tjhung \emph{et~al.}(2012)Tjhung, Marenduzzo, and Cates]{TjhungPNAS2012}
E.~Tjhung, D.~Marenduzzo and M.~E. Cates, \emph{Proc. Natl. Acad. Sci. USA}, 2012, \textbf{109}, 12381--12386\relax
\mciteBstWouldAddEndPuncttrue
\mciteSetBstMidEndSepPunct{\mcitedefaultmidpunct}
{\mcitedefaultendpunct}{\mcitedefaultseppunct}\relax
\EndOfBibitem
\bibitem[{De Magistris} \emph{et~al.}(2014){De Magistris}, Tiribocchi, Whitfield, Hawkins, Cates, and Marenduzzo]{DeMagistrisSM2014}
G.~{De Magistris}, A.~Tiribocchi, C.~A. Whitfield, R.~J. Hawkins, M.~E. Cates and D.~Marenduzzo, \emph{Soft Matter}, 2014, \textbf{10}, 7826--7837\relax
\mciteBstWouldAddEndPuncttrue
\mciteSetBstMidEndSepPunct{\mcitedefaultmidpunct}
{\mcitedefaultendpunct}{\mcitedefaultseppunct}\relax
\EndOfBibitem
\bibitem[Granek \emph{et~al.}(2022)Granek, Kafri, and Tailleur]{GranekPRL2022}
O.~Granek, Y.~Kafri and J.~Tailleur, \emph{Phys. Rev. Lett.}, 2022, \textbf{129}, 038001\relax
\mciteBstWouldAddEndPuncttrue
\mciteSetBstMidEndSepPunct{\mcitedefaultmidpunct}
{\mcitedefaultendpunct}{\mcitedefaultseppunct}\relax
\EndOfBibitem
\bibitem[Kim \emph{et~al.}(2024)Kim, Choe, and Baek]{KimPRE2024}
K.-W. Kim, Y.~Choe and Y.~Baek, \emph{Phys. Rev. E}, 2024, \textbf{109}, 014614\relax
\mciteBstWouldAddEndPuncttrue
\mciteSetBstMidEndSepPunct{\mcitedefaultmidpunct}
{\mcitedefaultendpunct}{\mcitedefaultseppunct}\relax
\EndOfBibitem
\bibitem[{Wang, Chao} \emph{et~al.}(2024){Wang, Chao}, {Lian, Wenchao}, {Li, Huishu}, {Tian, Wende}, and {Chen, Kang}]{WangNSO2024}
{Wang, Chao}, {Lian, Wenchao}, {Li, Huishu}, {Tian, Wende} and {Chen, Kang}, \emph{Natl. Sci. Open}, 2024, \textbf{3}, 20230066\relax
\mciteBstWouldAddEndPuncttrue
\mciteSetBstMidEndSepPunct{\mcitedefaultmidpunct}
{\mcitedefaultendpunct}{\mcitedefaultseppunct}\relax
\EndOfBibitem
\bibitem[Pei and Maes(2025)]{PeiPRE2025}
J.-H. Pei and C.~Maes, \emph{Phys. Rev. E}, 2025, \textbf{111}, L032101\relax
\mciteBstWouldAddEndPuncttrue
\mciteSetBstMidEndSepPunct{\mcitedefaultmidpunct}
{\mcitedefaultendpunct}{\mcitedefaultseppunct}\relax
\EndOfBibitem
\bibitem[Nikola \emph{et~al.}(2016)Nikola, Solon, Kafri, Kardar, Tailleur, and Voituriez]{NikolaPRL2016}
N.~Nikola, A.~P. Solon, Y.~Kafri, M.~Kardar, J.~Tailleur and R.~Voituriez, \emph{Phys. Rev. Lett.}, 2016, \textbf{117}, 098001\relax
\mciteBstWouldAddEndPuncttrue
\mciteSetBstMidEndSepPunct{\mcitedefaultmidpunct}
{\mcitedefaultendpunct}{\mcitedefaultseppunct}\relax
\EndOfBibitem
\bibitem[Shin \emph{et~al.}(2017)Shin, Cherstvy, Kim, and Zaburdaev]{ShinPCCP2017}
J.~Shin, A.~G. Cherstvy, W.~K. Kim and V.~Zaburdaev, \emph{Phys. Chem. Chem. Phys.}, 2017, \textbf{19}, 18338--18347\relax
\mciteBstWouldAddEndPuncttrue
\mciteSetBstMidEndSepPunct{\mcitedefaultmidpunct}
{\mcitedefaultendpunct}{\mcitedefaultseppunct}\relax
\EndOfBibitem
\bibitem[Li \emph{et~al.}(2017)Li, Wang, Tian, Ma, Xu, Zheng, and Chen]{LiSM2017}
H.-s. Li, C.~Wang, W.-d. Tian, Y.-q. Ma, C.~Xu, N.~Zheng and K.~Chen, \emph{Soft Matter}, 2017, \textbf{13}, 8031--8038\relax
\mciteBstWouldAddEndPuncttrue
\mciteSetBstMidEndSepPunct{\mcitedefaultmidpunct}
{\mcitedefaultendpunct}{\mcitedefaultseppunct}\relax
\EndOfBibitem
\bibitem[Shafiei~Aporvari \emph{et~al.}(2020)Shafiei~Aporvari, Utkur, Saritas, Volpe, and Stenhammar]{AporvariSoftMatter2020}
M.~Shafiei~Aporvari, M.~Utkur, E.~U. Saritas, G.~Volpe and J.~Stenhammar, \emph{Soft Matter}, 2020, \textbf{16}, 5609--5614\relax
\mciteBstWouldAddEndPuncttrue
\mciteSetBstMidEndSepPunct{\mcitedefaultmidpunct}
{\mcitedefaultendpunct}{\mcitedefaultseppunct}\relax
\EndOfBibitem
\bibitem[Melby \emph{et~al.}(2005)Melby, Reyes, Prevost, Robertson, Kumar, Egolf, and Urbach]{melby2005dynamics}
P.~Melby, F.~V. Reyes, A.~Prevost, R.~Robertson, P.~Kumar, D.~A. Egolf and J.~S. Urbach, \emph{J. Phys.: Condens. Matter}, 2005, \textbf{17}, S2689\relax
\mciteBstWouldAddEndPuncttrue
\mciteSetBstMidEndSepPunct{\mcitedefaultmidpunct}
{\mcitedefaultendpunct}{\mcitedefaultseppunct}\relax
\EndOfBibitem
\bibitem[Yamada \emph{et~al.}(2003)Yamada, Hondou, and Sano]{yamada2003coherent}
D.~Yamada, T.~Hondou and M.~Sano, \emph{Phys. Rev. E}, 2003, \textbf{67}, 040301\relax
\mciteBstWouldAddEndPuncttrue
\mciteSetBstMidEndSepPunct{\mcitedefaultmidpunct}
{\mcitedefaultendpunct}{\mcitedefaultseppunct}\relax
\EndOfBibitem
\bibitem[Kudrolli \emph{et~al.}(2008)Kudrolli, Lumay, Volfson, and Tsimring]{kudrolli2008swarming}
A.~Kudrolli, G.~Lumay, D.~Volfson and L.~S. Tsimring, \emph{Phys. Rev. Lett.}, 2008, \textbf{100}, 058001\relax
\mciteBstWouldAddEndPuncttrue
\mciteSetBstMidEndSepPunct{\mcitedefaultmidpunct}
{\mcitedefaultendpunct}{\mcitedefaultseppunct}\relax
\EndOfBibitem
\bibitem[Deseigne \emph{et~al.}(2010)Deseigne, Dauchot, and Chat{\'e}]{deseigne2010collective}
J.~Deseigne, O.~Dauchot and H.~Chat{\'e}, \emph{Phys. Rev. Lett.}, 2010, \textbf{105}, 098001\relax
\mciteBstWouldAddEndPuncttrue
\mciteSetBstMidEndSepPunct{\mcitedefaultmidpunct}
{\mcitedefaultendpunct}{\mcitedefaultseppunct}\relax
\EndOfBibitem
\bibitem[Deseigne \emph{et~al.}(2012)Deseigne, L{\'e}onard, Dauchot, and Chat{\'e}]{deseigne2012vibrated}
J.~Deseigne, S.~L{\'e}onard, O.~Dauchot and H.~Chat{\'e}, \emph{Soft Matter}, 2012, \textbf{8}, 5629--5639\relax
\mciteBstWouldAddEndPuncttrue
\mciteSetBstMidEndSepPunct{\mcitedefaultmidpunct}
{\mcitedefaultendpunct}{\mcitedefaultseppunct}\relax
\EndOfBibitem
\bibitem[Weber \emph{et~al.}(2013)Weber, Hanke, Deseigne, L{\'e}onard, Dauchot, Frey, and Chat{\'e}]{weber2013long}
C.~A. Weber, T.~Hanke, J.~Deseigne, S.~L{\'e}onard, O.~Dauchot, E.~Frey and H.~Chat{\'e}, \emph{Phys. Rev. Lett.}, 2013, \textbf{110}, 208001\relax
\mciteBstWouldAddEndPuncttrue
\mciteSetBstMidEndSepPunct{\mcitedefaultmidpunct}
{\mcitedefaultendpunct}{\mcitedefaultseppunct}\relax
\EndOfBibitem
\bibitem[Briand \emph{et~al.}(2018)Briand, Schindler, and Dauchot]{briand2018spontaneously}
G.~Briand, M.~Schindler and O.~Dauchot, \emph{Phys. Rev. Lett.}, 2018, \textbf{120}, 208001\relax
\mciteBstWouldAddEndPuncttrue
\mciteSetBstMidEndSepPunct{\mcitedefaultmidpunct}
{\mcitedefaultendpunct}{\mcitedefaultseppunct}\relax
\EndOfBibitem
\bibitem[Scholz \emph{et~al.}(2018)Scholz, Engel, and P{\"o}schel]{scholz2018rotating}
C.~Scholz, M.~Engel and T.~P{\"o}schel, \emph{Nat. Commun.}, 2018, \textbf{9}, 931\relax
\mciteBstWouldAddEndPuncttrue
\mciteSetBstMidEndSepPunct{\mcitedefaultmidpunct}
{\mcitedefaultendpunct}{\mcitedefaultseppunct}\relax
\EndOfBibitem
\bibitem[Digregorio \emph{et~al.}(2022)Digregorio, Levis, Cugliandolo, Gonnella, and Pagonabarraga]{digregorio2022unified}
P.~Digregorio, D.~Levis, L.~F. Cugliandolo, G.~Gonnella and I.~Pagonabarraga, \emph{Soft Matter}, 2022, \textbf{18}, 566--591\relax
\mciteBstWouldAddEndPuncttrue
\mciteSetBstMidEndSepPunct{\mcitedefaultmidpunct}
{\mcitedefaultendpunct}{\mcitedefaultseppunct}\relax
\EndOfBibitem
\bibitem[Ni \emph{et~al.}(2013)Ni, Stuart, and Dijkstra]{ni2013pushing}
R.~Ni, M.~A.~C. Stuart and M.~Dijkstra, \emph{Nat. Commun.}, 2013, \textbf{4}, 2704\relax
\mciteBstWouldAddEndPuncttrue
\mciteSetBstMidEndSepPunct{\mcitedefaultmidpunct}
{\mcitedefaultendpunct}{\mcitedefaultseppunct}\relax
\EndOfBibitem
\bibitem[K{\"u}mmel \emph{et~al.}(2015)K{\"u}mmel, Shabestari, Lozano, Volpe, and Bechinger]{kummel2015formation}
F.~K{\"u}mmel, P.~Shabestari, C.~Lozano, G.~Volpe and C.~Bechinger, \emph{Soft matter}, 2015, \textbf{11}, 6187--6191\relax
\mciteBstWouldAddEndPuncttrue
\mciteSetBstMidEndSepPunct{\mcitedefaultmidpunct}
{\mcitedefaultendpunct}{\mcitedefaultseppunct}\relax
\EndOfBibitem
\bibitem[van~der Meer \emph{et~al.}(2016)van~der Meer, Filion, and Dijkstra]{van2016fabricating}
B.~van~der Meer, L.~Filion and M.~Dijkstra, \emph{Soft Matter}, 2016, \textbf{12}, 3406--3411\relax
\mciteBstWouldAddEndPuncttrue
\mciteSetBstMidEndSepPunct{\mcitedefaultmidpunct}
{\mcitedefaultendpunct}{\mcitedefaultseppunct}\relax
\EndOfBibitem
\bibitem[Kumar \emph{et~al.}(2014)Kumar, Soni, Ramaswamy, and Sood]{kumar2014flocking}
N.~Kumar, H.~Soni, S.~Ramaswamy and A.~Sood, \emph{Nat. Commun.}, 2014, \textbf{5}, 4688\relax
\mciteBstWouldAddEndPuncttrue
\mciteSetBstMidEndSepPunct{\mcitedefaultmidpunct}
{\mcitedefaultendpunct}{\mcitedefaultseppunct}\relax
\EndOfBibitem
\bibitem[Soni \emph{et~al.}(2020)Soni, Kumar, Nambisan, Gupta, Sood, and Ramaswamy]{soni2020phases}
H.~Soni, N.~Kumar, J.~Nambisan, R.~K. Gupta, A.~Sood and S.~Ramaswamy, \emph{Soft Matter}, 2020, \textbf{16}, 7210--7221\relax
\mciteBstWouldAddEndPuncttrue
\mciteSetBstMidEndSepPunct{\mcitedefaultmidpunct}
{\mcitedefaultendpunct}{\mcitedefaultseppunct}\relax
\EndOfBibitem
\bibitem[Son \emph{et~al.}(2024)Son, Choe, Kwon, Rigon, Baek, and Kim]{son2024dynamics}
K.~Son, Y.~Choe, E.~Kwon, L.~G. Rigon, Y.~Baek and H.-Y. Kim, \emph{Soft Matter}, 2024, \textbf{20}, 2777--2788\relax
\mciteBstWouldAddEndPuncttrue
\mciteSetBstMidEndSepPunct{\mcitedefaultmidpunct}
{\mcitedefaultendpunct}{\mcitedefaultseppunct}\relax
\EndOfBibitem
\bibitem[Elgeti and Gompper(2009)]{elgeti2009self}
J.~Elgeti and G.~Gompper, \emph{Europhys. Lett.}, 2009, \textbf{85}, 38002\relax
\mciteBstWouldAddEndPuncttrue
\mciteSetBstMidEndSepPunct{\mcitedefaultmidpunct}
{\mcitedefaultendpunct}{\mcitedefaultseppunct}\relax
\EndOfBibitem
\bibitem[Deblais \emph{et~al.}(2018)Deblais, Barois, Guerin, Delville, Vaudaine, Lintuvuori, Boudet, Baret, and Kellay]{deblais2018boundaries}
A.~Deblais, T.~Barois, T.~Guerin, P.-H. Delville, R.~Vaudaine, J.~S. Lintuvuori, J.-F. Boudet, J.-C. Baret and H.~Kellay, \emph{Phys. Rev. Lett.}, 2018, \textbf{120}, 188002\relax
\mciteBstWouldAddEndPuncttrue
\mciteSetBstMidEndSepPunct{\mcitedefaultmidpunct}
{\mcitedefaultendpunct}{\mcitedefaultseppunct}\relax
\EndOfBibitem
\bibitem[Wensink and L{\"o}wen(2008)]{wensink2008aggregation}
H.~Wensink and H.~L{\"o}wen, \emph{Phys. Rev. E}, 2008, \textbf{78}, 031409\relax
\mciteBstWouldAddEndPuncttrue
\mciteSetBstMidEndSepPunct{\mcitedefaultmidpunct}
{\mcitedefaultendpunct}{\mcitedefaultseppunct}\relax
\EndOfBibitem
\bibitem[Boudet \emph{et~al.}(2021)Boudet, Lintuvuori, Lacouture, Barois, Deblais, Xie, Cassagnere, Tregon, Br{\"u}ckner, Baret,\emph{et~al.}]{boudet2021collections}
J.-F. Boudet, J.~Lintuvuori, C.~Lacouture, T.~Barois, A.~Deblais, K.~Xie, S.~Cassagnere, B.~Tregon, D.~B. Br{\"u}ckner, J.-C. Baret \emph{et~al.}, \emph{Sci. Rob.}, 2021, \textbf{6}, eabd0272\relax
\mciteBstWouldAddEndPuncttrue
\mciteSetBstMidEndSepPunct{\mcitedefaultmidpunct}
{\mcitedefaultendpunct}{\mcitedefaultseppunct}\relax
\EndOfBibitem
\bibitem[Lee \emph{et~al.}(2023)Lee, Sch{\"o}nh{\"o}fer, and Glotzer]{lee2023complex}
S.~Y. Lee, P.~W. Sch{\"o}nh{\"o}fer and S.~C. Glotzer, \emph{Sci. Rep.}, 2023, \textbf{13}, 22773\relax
\mciteBstWouldAddEndPuncttrue
\mciteSetBstMidEndSepPunct{\mcitedefaultmidpunct}
{\mcitedefaultendpunct}{\mcitedefaultseppunct}\relax
\EndOfBibitem
\end{mcitethebibliography}
\bibliographystyle{rsc} 

\end{document}